\documentclass[iop,apj,numberappendix,appendixfloats]{emulateapj}
\usepackage{graphicx}
\pdfoutput=1

\LongTables
\def \lyaf {Ly$\alpha$ forest}

\def \lya  {Ly$\alpha$}

\def \lyb  {Ly$\beta$}

\def \ly5  {Ly-5}
\def \ly6  {Ly-6}
\def \ly7  {Ly-7}

\def \lyaf {Lyman--$\alpha$ forest}

\newcommand{\kms}{${\rm km\,s}^{-1}$}

\def\lesssim{\mathrel{\hbox{\rlap{\hbox{%
 \lower4pt\hbox{$\sim$}}}\hbox{$<$}}}}
\def\gtrsim{\mathrel{\hbox{\rlap{\hbox{%
 \lower4pt\hbox{$\sim$}}}\hbox{$>$}}}}
                
\let\ga=\gtrsim

\newcommand{\hi}{H$\;${\small\rm I}\relax}

\newcommand{\civ}{C$\;${\small\rm IV}\relax}

\newcommand{\ovi}{O$\;${\small\rm VI}\relax}

\newcommand{\mgii}{Mg$\;${\small\rm II}\relax}

\newcommand{\civt}{C$\;${\scriptsize\rm IV}\relax}

\newcommand{\ovit}{O$\;${\scriptsize\rm VI}\relax}

\def \nqsos {170}
\def \nspectra {247}
\def \ntotspectra {567}

\submitted{Submitted to the Astronomical Journal}
\shortauthors{O'Meara et al.}
\shorttitle{KODIAQ DR1}

\begin{document}

\title{The First Data Release of the KODIAQ Survey}

\author{
J.M. O'Meara\altaffilmark{1},
N. Lehner\altaffilmark{2},
J.C. Howk\altaffilmark{2},
J.X. Prochaska\altaffilmark{3},
A.J. Fox\altaffilmark{4},
M. A. Swain\altaffilmark{5},
C. R. Gelino\altaffilmark{6},
G. B. Berriman\altaffilmark{6},
\& H. Tran\altaffilmark{7}
}
\altaffiltext{1}{Department of Chemistry and Physics, Saint Michael's College, One Winooski Park, Colchester, VT 05439}
\altaffiltext{2}{Department of Physics, University of Notre Dame, 225 Nieuwland Science Hall, Notre Dame, IN 46556}
\altaffiltext{3}{University of California/Lick Observatory, Santa Cruz, 1156 High Street, CA 95064}
\altaffiltext{4}{Space Telescope Science Institute, Baltimore, MD 21218}
\altaffiltext{5}{Raytheon, 299 N Euclid Avenue, Suite 500, Pasadena, CA 91101}
\altaffiltext{6}{NASA Exoplanet Science Institute, Infrared Processing
  and Analysis Center, California Institute of Technoly, Pasadena, CA  91125}
\altaffiltext{7}{W. M. Keck Observatory, 65-1120 Mamalahoa Hwy, Kamuela, HI 96743}

\begin{abstract}
  We present and make publicly available the first data release (DR1)
  of the Keck Observatory Database of Ionized Absorption toward
  Quasars (KODIAQ) survey. The KODIAQ survey is aimed at studying
  galactic and circumgalactic gas in absorption at high-redshift, with
  a focus on highly-ionized gas traced by \ovit, using the HIRES
  spectrograph on the Keck-I telescope.  KODIAQ DR1 consists of a
  fully-reduced sample of \nqsos\ quasars at $0.29 < z_{\rm em} <
  5.29$ observed with HIRES at high resolution ($36,000 \le R \le
  103,000$) between 2004 and 2012.  DR1 contains \nspectra\ spectra
  available in continuum normalized form, representing a sum total
  exposure time of $\sim 1.6$ megaseconds.  These coadded spectra
  arise from a total of \ntotspectra\ individual exposures of quasars
  taken from the Keck Observatory Archive (KOA) in raw form and
  uniformly processed using a HIRES data reduction package made
  available through the XIDL distribution.  DR1 is publicly available
  to the community, housed as a higher level science product at the
  KOA.  We will provide future data releases that make further QSOs,
  including those with pre-2004 observations taken with the
  previous-generation HIRES detectors.
\end{abstract}


\section{Introduction}\label{intro}

The High Resolution Echelle Spectrometer (HIRES) \cite{vogt94} at the
Keck-I telescope on Maunakea has contributed significantly to our
knowledge of the intermediate- to high-redshift universe.  In
particular, observations of absorption lines toward background quasars
with HIRES have provided a number of fundamental discoveries and
studies.  HIRES observations have provided experimental tests of Big
Bang Nucleosynthesis and have help determine the cosmological baryon
density (e.g., \citealt{burles98a,omeara01,cooke14}).  They have
discovered pristine gas clouds \citep{fumagalli11} and determined the
metal budgets over 10 decades in \hi\ column density from the \lyaf\
\citep{simcoe04} to the damped Lyman alpha systems
\citep{wolfe05}. They have explored the galaxy-IGM connection
\citep{steidel10} and have constrained the thermal history of the
universe \citep{bolton14}. They have helped constrain dark matter
models \citep{viel13}, have provided critical tests to cosmological
simulations \citep{tytler07}, and been used to study changes in the
fine-structure constant with redshift \citep{murphy01,webb01,murphy03}.
Indeed, at the time of submission, HIRES appears in the title or
abstract of over 1000 papers totaling over 20,000 citations.


Remarkably, these successful studies have primarily been achieved with
single objects, or with samples numbering in the tens of sightlines.
This largely reflects the limited allocations of Principal
Investigators (PIs) with direct access to the Keck telescopes.  The
compilation of a significant portion of the quasar sightlines observed
with HIRES in the Keck Observatory Archive
(KOA\footnote{http://www2.keck.hawaii.edu/koa/public/koa.php}),
however, now facilitates the study of hundreds of quasar sightlines by
making the data available to the community in raw form.  In 2009, we
were awarded a NASA ADAP grant (PI: Lehner) to use this rich QSO
database ($> 600$ QSOs as of this writing) to study the hot ionized
\ovi\ gas in the circumgalactic medium of $2 < z < 4$ galaxies in a
sample of \hi-selected strong absorbers ($\log N_{\rm HI}\ga 17$,
i.e., the Lyman limit and damped Ly$\alpha$ systems).  The first
results from our Keck Observatory Database of Ionized Absorption
toward Quasars (KODIAQ) survey are presented in \citet{lehner14}.

As part of the KODIAQ survey, we have uniformly processed the HIRES
spectra of hundreds of QSOs. In this work, we present the first data
release (DR1) from the survey.  We make available to the community
continuum normalized 1-D spectra for \nqsos\ QSOs and make them
publicly available at the KOA.  This work joins and extends the public
data releases of Keck data from \cite{Xgrb}, \cite{Xesi}, \cite{Xdla},
and \cite{Rafelski}, and is complementary to the ESO Advanced Data
Products Quasar Sample (e.g., \citealt{zafar}).  Here we present the
details of our work to produce this archive. An outline of the paper
is as follows: In section \ref{data}, we describe the basic properties
of the data, it's reduction, and its availability at the KOA.  In
section \ref{dr1}, we describe the specific properties of the DR1
sample.  Finally, in section \ref{future} we describe future planned
data releases and summarize.

\section{The Data}\label{data}
Data taken over the last two decades from the HIRES spectrograph fall
into three general categories.  The first is the earliest data from
1995--1997, where the light is dispersed onto a single Tektronix
$2048\times 2048$ chip (see \citealt{vogt94} for details).  The second
category spans the years 1997--2004, when HIRES could be configured
using either a red or blue sensitive cross-disperser to enhance
throughput over specific wavelength ranges, but still feeding a single
CCD.  Finally, the third category, covering data from 2004 to the
present, has the original single CCD replaced by a 3 chip mosaic, with
dramatically increased blue sensitivity.  The data in DR1 are all
drawn from this last category.

HIRES has a number of user-selectable instrument element settings.
These include detector binning, entrance slit width and length (set
simultaneously by choosing a decker), and
echelle and cross-disperser angle.  The latter two elements determine
the wavelength coverage for a given setup.  For data taken
with the blue cross-disperser, the data frequently have continuous
wavelength coverage (with some overlaps between echelle orders),
albeit with two small gaps in coverage due to the spacing between the
chips in the mosaic.  Data obtained with the red cross disperser often
has additional gaps in wavelength coverage, due to the increasing
spatial extent of the dispersed light becoming larger than the extent
of the chip. The slit width sets the resolution of the data.  Table~1
gives the spectral resolution associated with the HIRES deckers
 that span the DR1 sample.  The vast majority of HIRES data in DR1
 were obtained with the C1 or C5 decker that provide FWHM
$R=48,000$ and $R=36,000$, corresponding to velocity resolutions of
6.2 and 8.3 \kms\ (FWHM), respectively.

\subsection{Raw Data}\label{rawdata}
The data for the KODIAQ survey are taken entirely from the KOA, i.e.,
they constitute a publicly-available sample.  While a quantity of
HIRES observations of quasar absorption lines still remain
unavailable,\footnote{Some PIs have requested and received proprietary
  times exceeding the nominal 1.5\,years} the KOA sample constitutes a
high percentage of all quasars observed to date with HIRES, as
measured by comparing the data in the sample with the Keck-I telescope
schedule over the last two decades.  Through searching by a
combination of Principal Investigator (PI) and project titles
associated with HIRES data at the KOA, we downloaded all the QSO data
in raw form for reduction and analysis.  Including calibration files,
the sample contains many tens of thousands of individual HIRES
exposures totaling $\sim$250 Gb.  The data were grouped by observing
run, where a single PI would observe with HIRES for a single night, or
over multiple nights close in time.  The choice of observing run
grouping was made to best facilitate data reduction of objects with
nearly identical HIRES configurations.

\subsection{Data Reduction}\label{redux}
Once grouped into observing runs, the data were uniformly reduced
using the HIRedux
code\footnote{http://www.ucolick.org/$\sim$xavier/HIRedux/} developed as
part of the
XIDL\footnote{http://www.ucolick.org/$\sim$xavier/IDL/index.html} suite of
astronomical routines in IDL.  HIRedux was developed as a variant 
of the reduction pipeline\footnote{http://web.mit.edu/$\sim$burles/www/MIKE/} 
for the MIKE spectrograph on the Magellan
Clay telescope at Las Campanas Observatory which further evolved in
the MaSE pipeline \citep{mase}.  

Briefly, the HIRedux workflow is as follows: First, the raw data are
read into the pipeline and sorted into ``setups'', where each setup is
a unique combination of decker, binning, and dispersing element choice
and configuration. Next, for each setup, a bias level and gain is
determined from the data, and a flat field image is constructed by a
median combine of individual exposures.  The flat field is used to
determine the echelle order edges via polynomial fitting, and to
determine the illumination of light along the slit (i.e., the tilt),
which is not constant.  Traditional pixel-to-pixel variation
flat-fielding is performed using archival pixel flat fields obtained
every few semesters by JXP.  These were acquired with a clever recipe
kindly suggested by R. Kibrick: a series of spectral images obtained
with the cover to the cross-disperser in place.  This cover is
partially reflective and yields a relatively uniform image on one
detector of the CCD mosaic when the cross-disperser grating is tilted
to $-3.1, -4.1$ or $-5.5$ for the blue, green and red detectors,
respectively.  The resultant, normalized image corrects for
small-scale variations in the detector (e.g., defects, ``pocks'',
etc.) that challenge a traditional approach to flat fielding.

With the orders spatially defined, the next step is to
determine a two-dimensional wavelength image for the chip by first
extracting down the center of each order an exposure of a ThAr arc
lamp.  This extracted arc spectrum is then compared to a suite of
archival ThAr databases and the algorithm solves for a two-dimensional 
wavelength solution rejecting outliers or spurious
arc line features.   The residuals to the arc line solutions
are generally at the sub-pixel level.  If multiple exposures of a science target
exist, cosmic rays are identified by analyzing the ratio of each
exposure to the median combined frame in standard ways.

Next, the data are sky-subtracted and the
object extracted on an order-by-order basis.  The specifics of the sky
subtration and object extraction are given in \citet{mase} who
describes the MaSE extraction pipeline (used to process data from the
Magellan MagE spectrograph), which shares a common codebase
with HIRedux.  The output of the extraction is an
order-by-order spectrum of counts versus vacuum wavelength.  For
multiple exposures of a science target with the same setup, the
individual orders are weighted-mean combined, with the weights coming
from the signal to noise of the individual exposures.

\subsection{Continuum Placement}\label{continuum}
For most types of analysis, quasar absorption line data are required
to be in continuum normalized form.  To put the data in this form, the
shape of both the spectrograph blaze function and the underlying
quasar SED must be removed via the process of continuum-fitting.  Many
spectrographs are able, via the use of a spectrophotometric standard
star, to remove the instrumenal response and flux calibrate the data
to aid in this process. Unfortunately, HIRES can not be reliably flux
calibrated (e.g., \citealt{suzuki}).  As such, the data are fitted by
hand, one echelle order at a time.

\begin{figure*}[ht]
\plotone{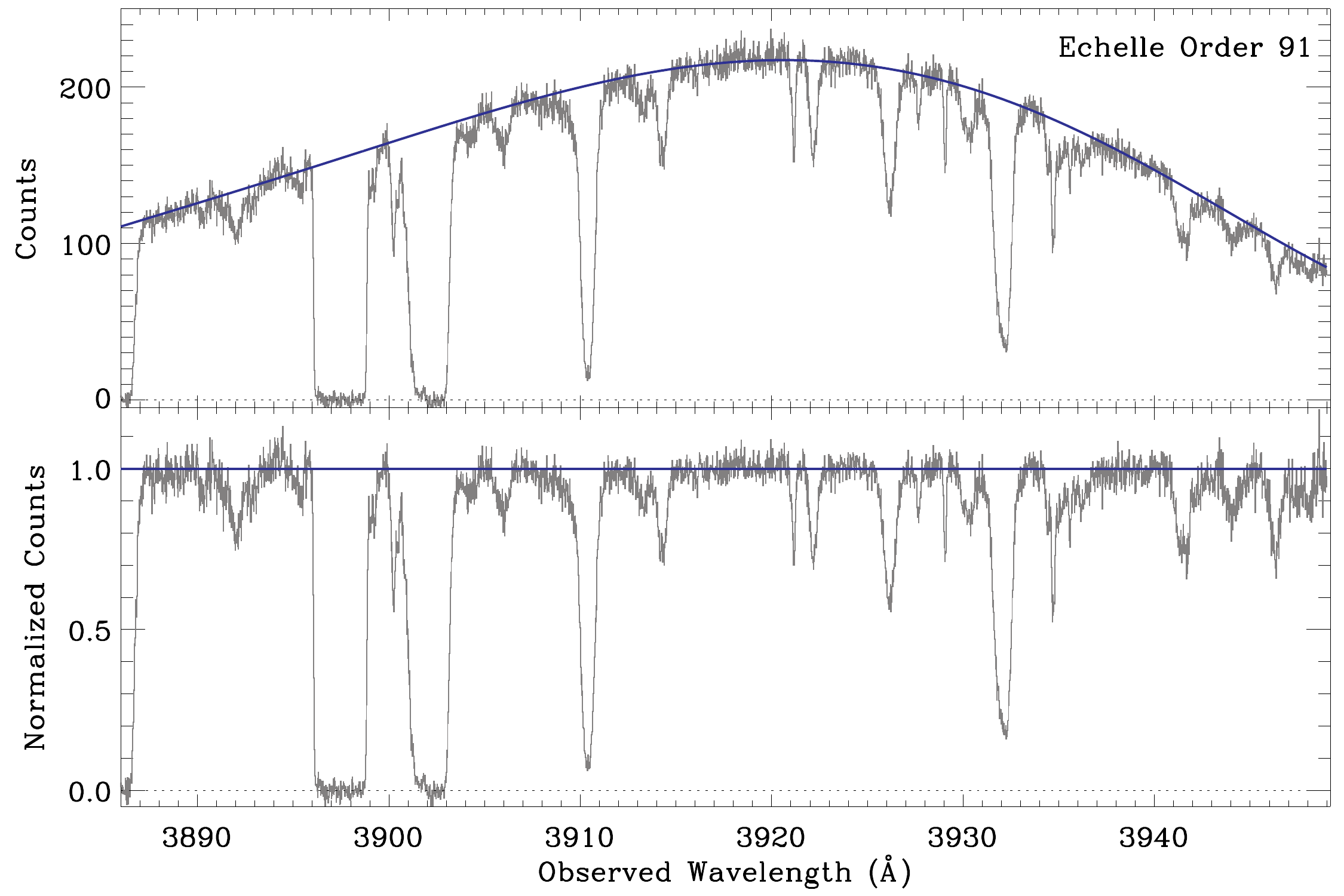}  
\caption{Example of a continuum fit to a single HIRES echelle order
  (Order 91) of the quasar J220852-194400 ($z_{\rm em} = 2.573$).  
The upper panel shows the output of the HIRedux
  extraction, and the adopted continuum level.  The lower panel shows
  the normalized echelle order.\label{fig_cont}}
\end{figure*}

To minimize variation in continuum-fitting methodologies, a single
member of the team (JMO) performed the continuum-fitting for the DR1
sample.  The exact method of fitting varies by quasar to quasar
depending on a combination of the signal to noise of the data, whether
the regions to be fit are blueward or redward of the quasar's \lya\
line (i.e., if the data covers the \lyaf), the redshift of the quasar
(larger redshift quasars have significantly more \lyaf\ absorption),
the shape of the blaze function in a given order, and the occurrence of
strong features such as quasar emission lines or damped Lyman alpha 
absorbers.  Nevertheless, a basic feature of all the continuum fits is
that they were performed using Legendre polynomial fits to each
spectral order, with the order of the Legendre polynomial ranging
nominally anywhere from 4 to 12 (or more in rare cases), 
depending on the factors given above.  In the
\lyaf , the continuum is anchored at points deemed by eye to be
absorption free. In regions redward of the \lyaf, significantly more
regions of the spectrum are absorption-free, and the fit is in general
more reliable and objective. Given
the number of spectra present in the DR1, the number of echelle
orders fit is well over 8000.  An example of a continuum fit to a
single HIRES echelle order is shown in Figure \ref{fig_cont}.

\begin{figure*}[ht]
\plotone{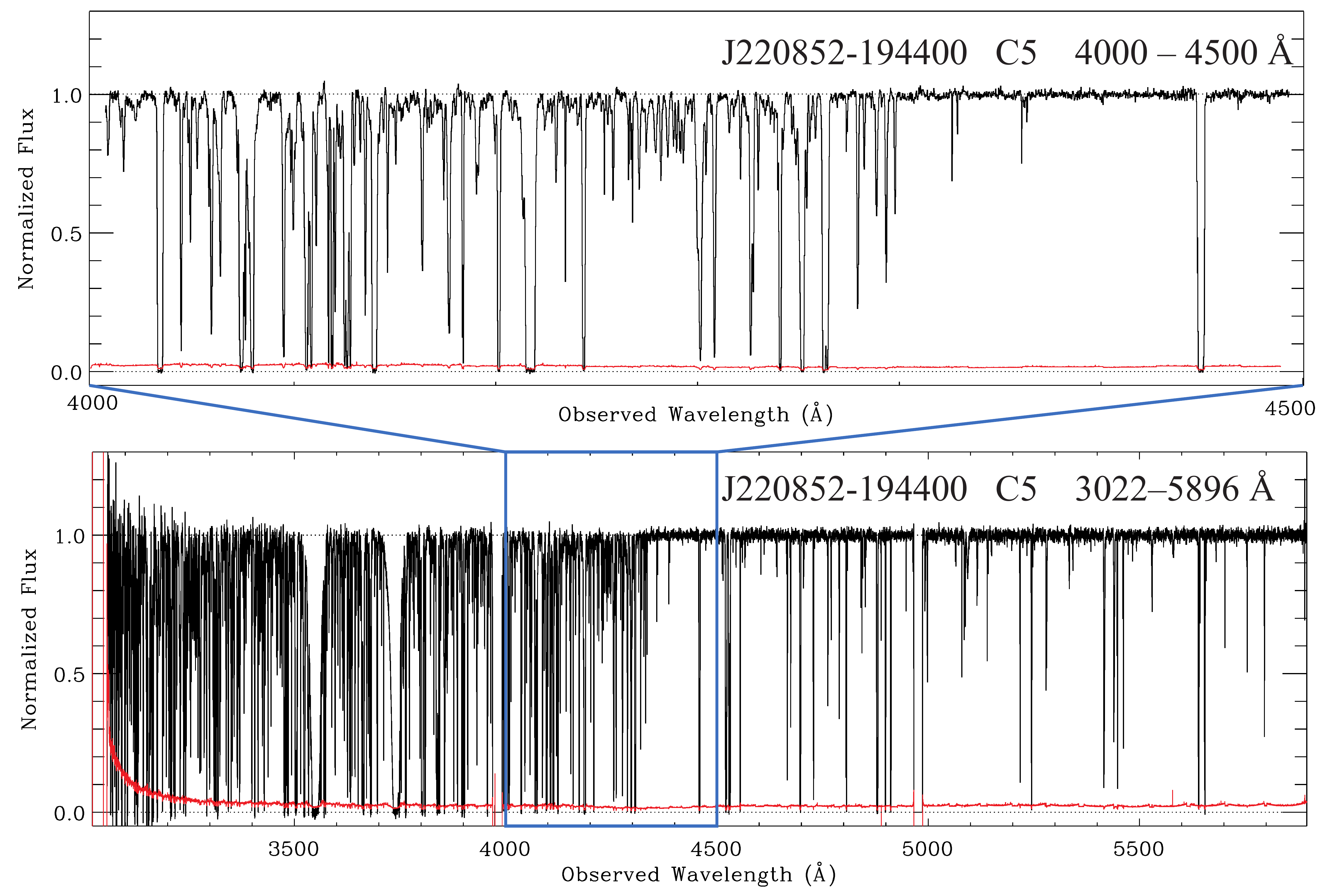}  
\caption{Lower panel: Example of a continuum normalized, coadded 1-D spectrum from
  DR1 (the same quasar is shown in Figure \ref{fig_cont}).  The
  1$\sigma$ error array is shown in red.  The quasar name, HIRES
  decker used, and the wavelength coverage of the data are listed in
  the figure. Upper panel:  A zoom in of the same spectrum, covering
  the wavelengths surrounding \lya\ at $z=z_{em}$.\label{fig_spectrum}}
\end{figure*}

Since the true quasar SED is unknown, especially at the resolution of
HIRES, an exact quantification of errors in the continuum fit cannot
be obtained.  Previous attempts at estimating continuum level errors
(e.g., \citealt{faucher-giguere}) typically find the error to be at the
few percent level or less in moderate to high signal to noise data
(e.g., S/N$\, \ga 10$), with the errors
increasing as either the signal to noise decreases, or the redshift
of the quasar increases.  As we will see in section \ref{dr1}, the DR1
sample covers a wide range of both quantities.  We also note that the
continuum placement will incur additional uncertainties in
regions within $\sim$$1000$ \kms\ of strong emission/absorption features
such as damped Lyman alpha absorbers, or BAL features in the quasar
spectrum itself.  We recommend that users of DR1 data either refine
our continuum estimate, or place their own continuum levels on the
extracted data if a more precise continuum estimate is demanded in their
analysis.  As we show below, the DR1 data are housed at KOA in a
manner such as to facilitate a re-assignment of the continuum level as
users see fit.  After the continuum level is assigned to each order,
the data are combined into separate one-dimensional flux and error
spectra, with regions of echelle order overlap in wavelength space
added together weighted by signal to noise.  An example of this final
product is shown in Figure \ref{fig_spectrum}.  The full HIRES header is
preserved from the raw data, along with KOA- and KODIAQ-specific header
cards appended to the end of the original data header. 

\subsection{KODIAQ at the KOA}\label{koa}
Beginning in May 2015, data from DR1 and all subsequent data releases from
KODIAQ will be available via the ``Contributed Datasets'' section of
the KOA.\footnote{https://koa.ipac.caltech.edu/Datasets}  Users can search for
individual quasar sightlines, or can download the entire set of
one-dimensional continuum normalized final science products. Each
quasar sightline contains a machine readable summary file detailing
all observations of the quasar, one-dimensional flux and error files
for each (coadded) observation in \texttt{FITS} format, and a preview image.  To
facilitate analysis according to the users needs, all intermediate
data reduction steps will be also available at the KOA, and can be
downloaded on an observing run by observing run basis.  The intermediate
data are grouped by observing run since many quasars can be observed in
a single HIRES setup on a given run. Each observing run has a file
structure dictated by HIRedux. In particular, users may wish to
download the intermediate data products to assign their own continuum
level to the individual HIRES orders.

\section{Properties of KODIAQ DR1}\label{dr1}
The KODIAQ DR1  comprises \nqsos\ quasars observed in the mosaic
configuration of HIRES between 2004 and 2012.\footnote{One quasar is a
  lensed system, so the number of individual sightlines is 171}  
The DR1 sample is summarized in Table 2.  The convention adopted here 
is to name each quasar after their J2000 RA/DEC as resolved by SIMBAD
or SDSS.  For example, the quasar with J2000 coordinates 
01:08:06.41+16:35:50.0 receives the name J010806+163550.  In the case
of very close quasar pairs or lensed quasars, the name is appended
with A,B, and so on.
Many of the quasars have been observed by multiple PIs, and/or with
multiple HIRES setups, such that the total number of spectra available
in DR1 is \nspectra\, representing coadditions from \ntotspectra\
individual HIRES observations.  Exposure times range from the hundreds
to tens of thousands of seconds, and the combined total exposure time
for DR1 is approximately 1.6 megaseconds. 

\subsection{General Properties}\label{general}
\begin{figure*}[ht]
\epsscale{0.7} 
\plotone{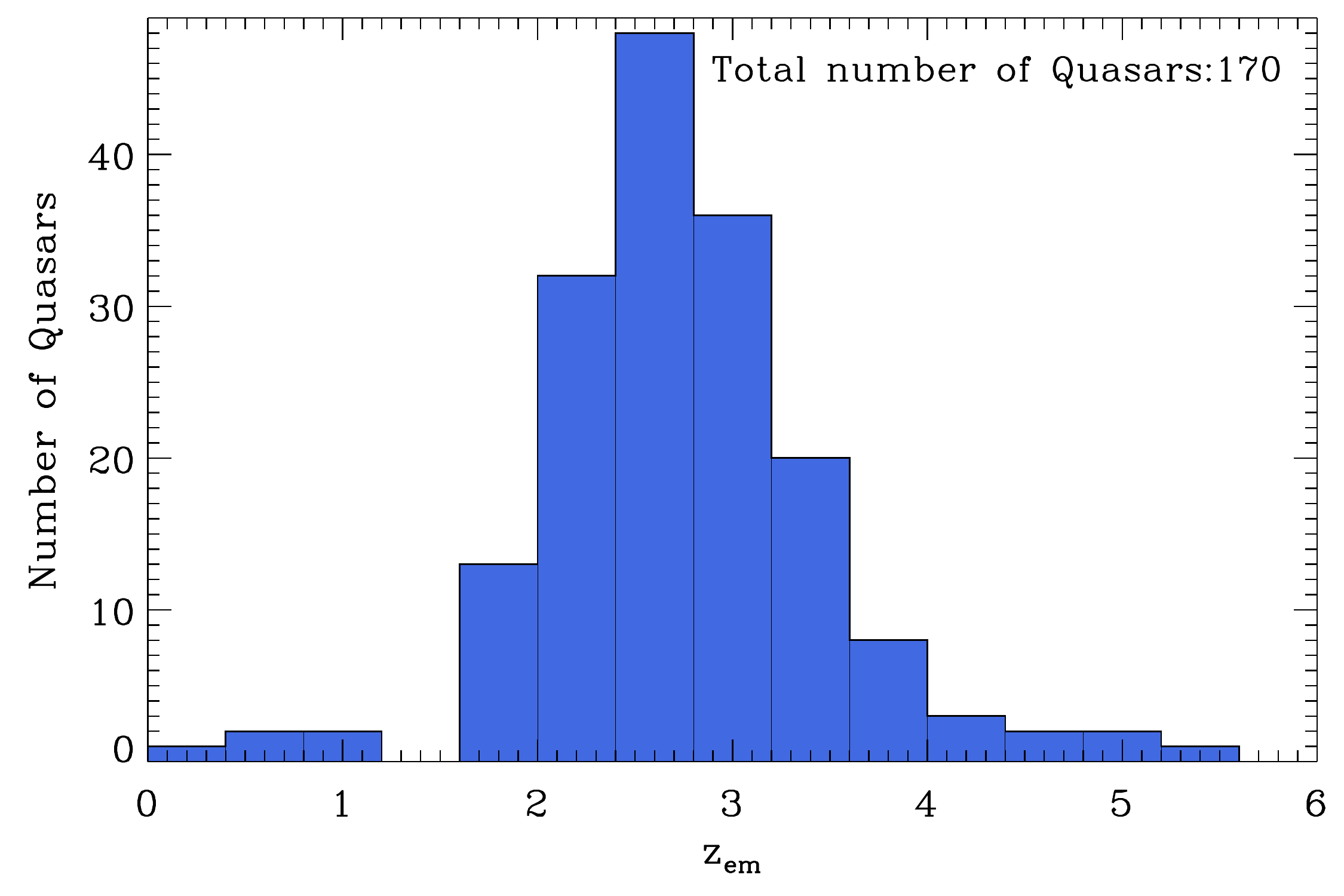}  
\caption{Redshift distribution of KODIAQ DR1 quasars. \label{fig_zhist}}
\end{figure*}

\begin{figure*}[ht]
\plotone{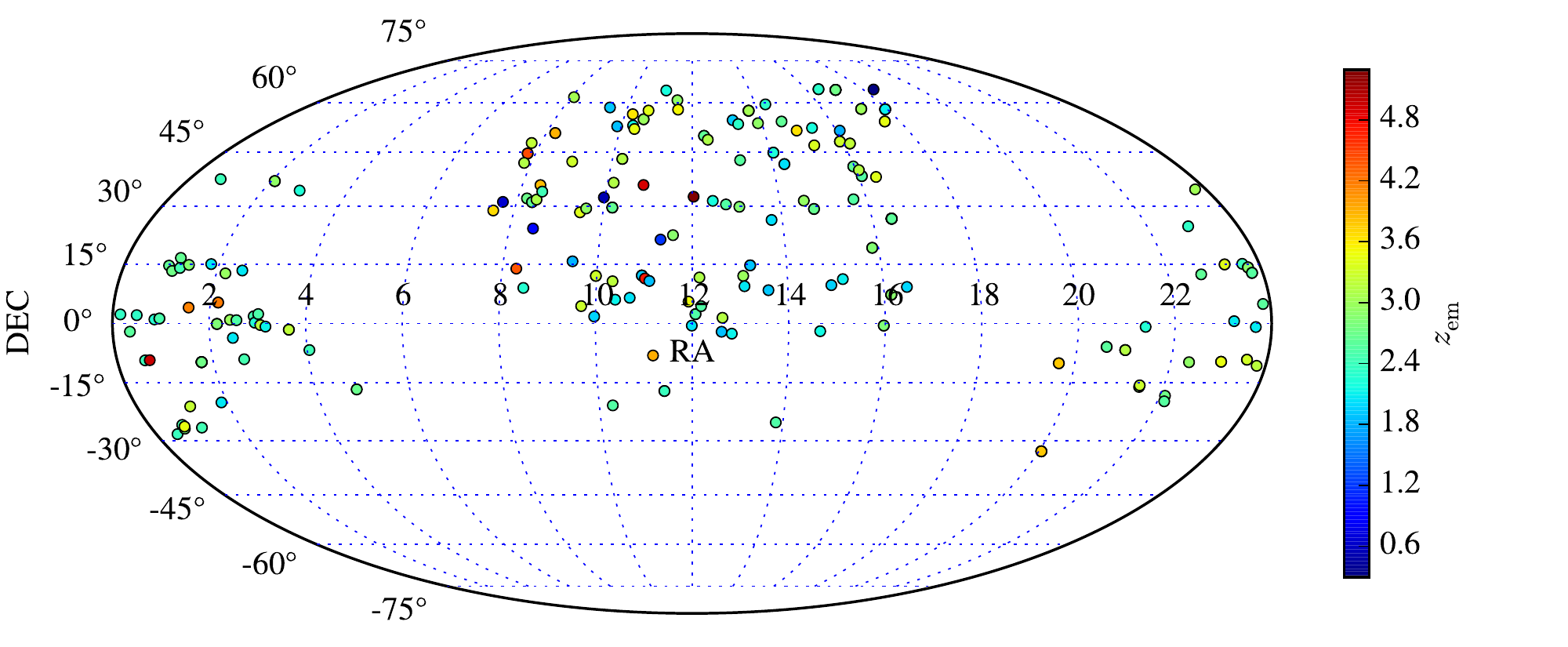}  
\caption{Aitoff sky distribution of the KODIAQ quasars in R.A. (hours)
  and Dec. (degrees). Each quasar is represented by a circle with its color encoding its emission redshift. The large regions on the sky without data stem from the Keck-I south pointing limits and the blocking on the sky of the Milky Way.\label{fig_map}}
\end{figure*}

\begin{figure*}[ht]
\epsscale{0.7} 
\plotone{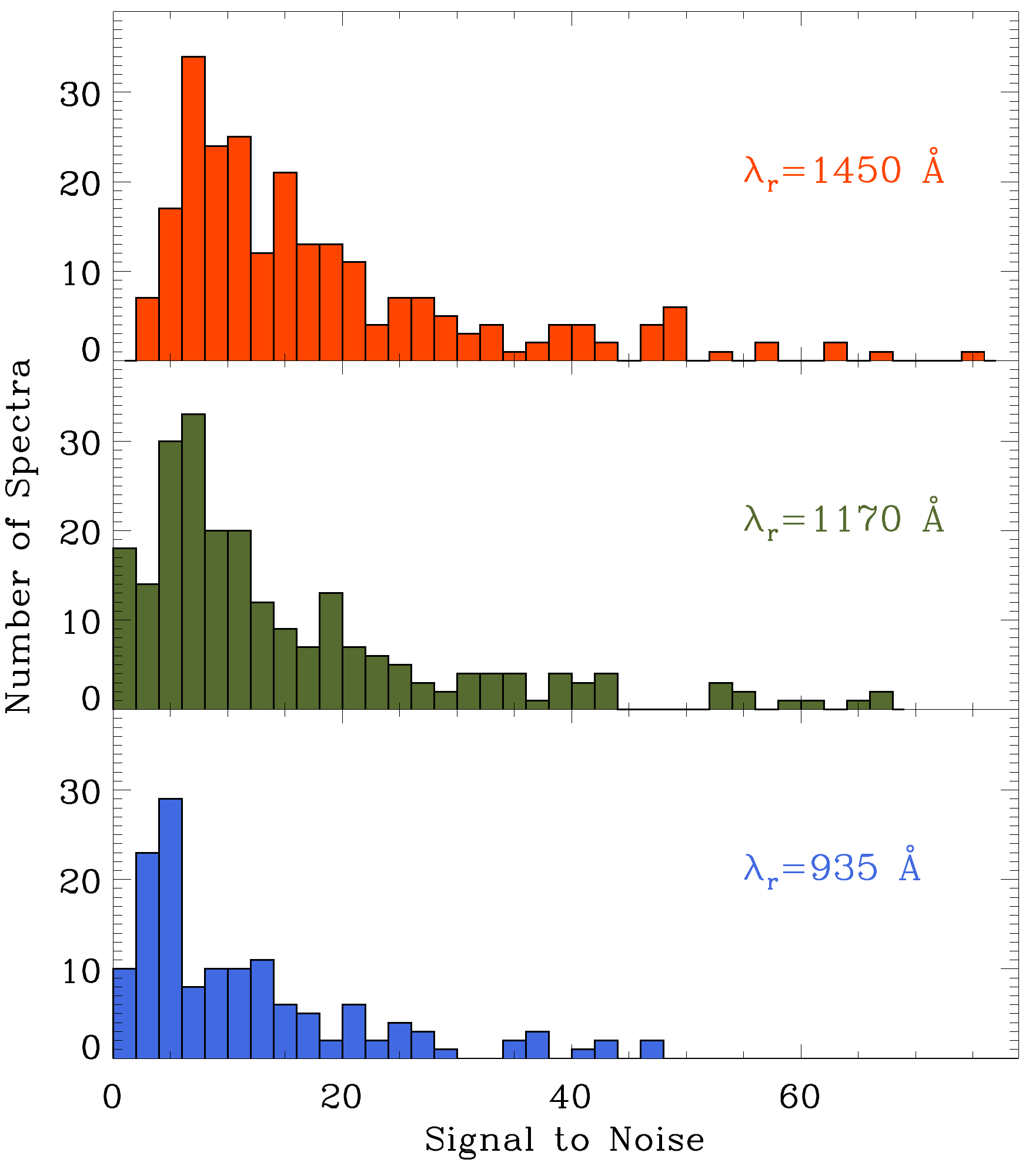}  
\caption{Median signal to noise per 2.6 \kms\ interval in a window $\pm 5$\AA\ about $\lambda =
  935,1170,1450 $ \AA\ for the \nspectra\ spectra in the KODIAQ sample.  \label{fig_snrhist}}
\end{figure*}

\begin{figure*}[ht]
\plotone{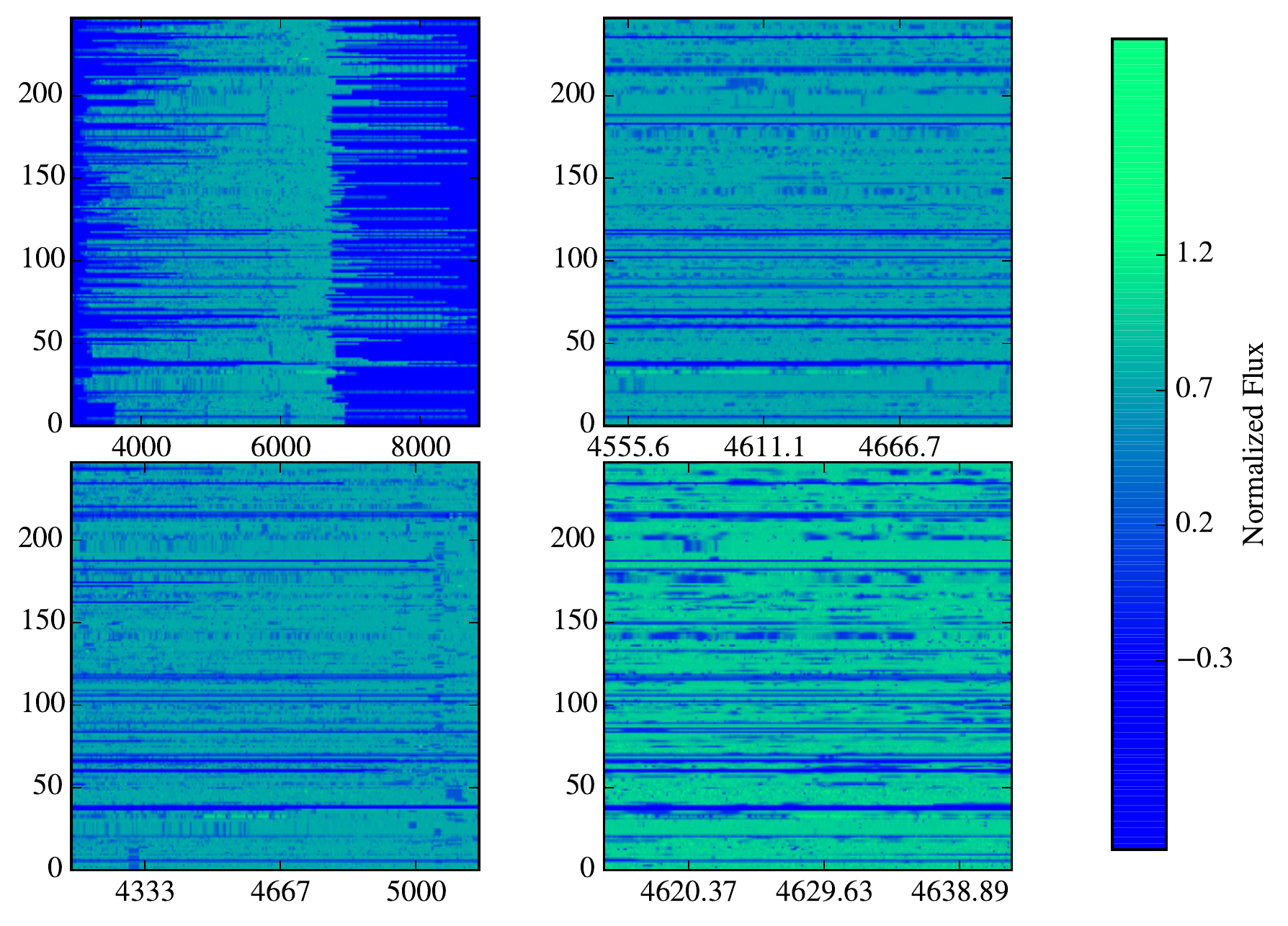}  
\caption{Spectral image of the full DR1 dataset.  The data are
  rebinned to 500 spectra pixels.  The panels show the dataset at
  various wavelength scales to illustrate the effects of instrument
  setup at the largest scales (upper-left panel) down the level of the 
the influence of the \lyaf\ (lower-right panel). \label{fig_spec_img}}
\end{figure*}

Figure \ref{fig_zhist} shows the quasar emission redshift distribution
for DR1, and shows that the DR1 data span a redshift range of $0.29 <
z_{\rm em} < 5.29$ in a roughly gaussian distribution, with a mean value
of $z_{\rm em}=2.725$  and a standard deviation of 0.738.   
The distribution of DR1 quasars on the sky is shown in Figure
\ref{fig_map}; these primarily cover the North Galactic Pole.
The data vary significantly in signal to noise.  In Figure
\ref{fig_snrhist}, we display the signal-to-noise S/N of the KODIAQ spectra in the
following way.  
For each of the  \nspectra\ spectra, we calculate the
median S/N per 2.6\kms\ interval (i.e., two unbinned HIRES pixels)
over a window of $\lambda = \pm 5$\AA\ on either side of 3 wavelength bins in the quasar
rest frame
These wavelengths are chosen to give a sense of the data near the
quasar Lyman limit, the regions encompassing the \lyaf\ and a region
sampling \lyaf-free wavelengths for metal line searches. Figure
\ref{fig_snrhist} illustrates that a significant fraction of the data
has S/N\,$>10$, facilitating a wide variety of future studies. 
For further illustration of the data, Figure\ref{fig_spec_img}
 shows the full spectral dataset rebinned to 500 spectral pixels
ordered from top-to-bottom by quasar
emission redshift.  The echelle order gaps and HIRES setups are
evident on large scales, as well as the
large-scale influence of the \lyaf\ on the smaller scales.

\subsection{Cosmological Properties}\label{cosmological}
\begin{figure*}[ht]
\epsscale{0.7} 
\plotone{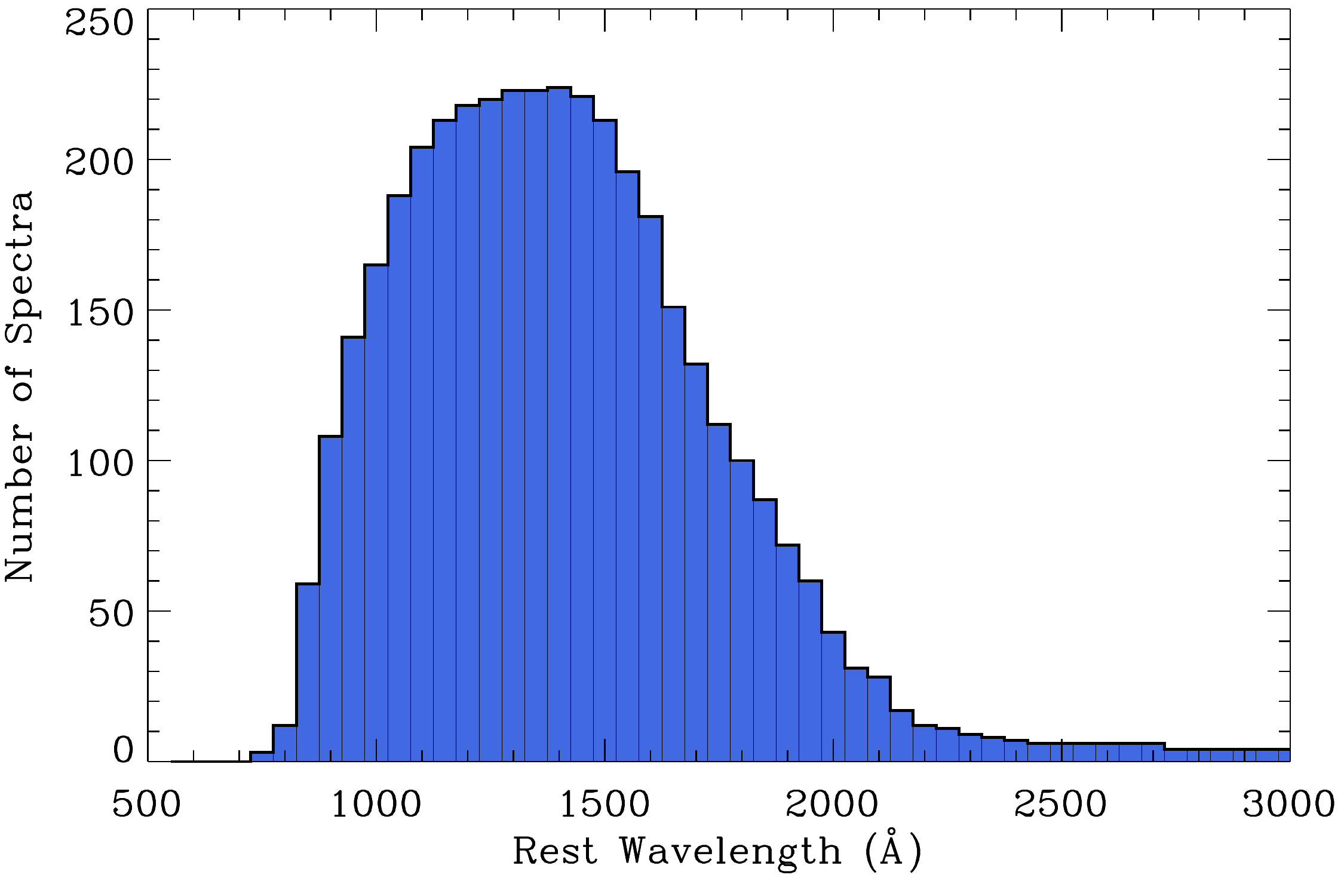}  
\caption{Rest frame wavelength spectral coverage of the \nspectra\
  spectra  in KODIAQ DR1. \label{fig_resthist}}
\end{figure*}

\begin{figure*}[ht]
\epsscale{0.7} 
\plotone{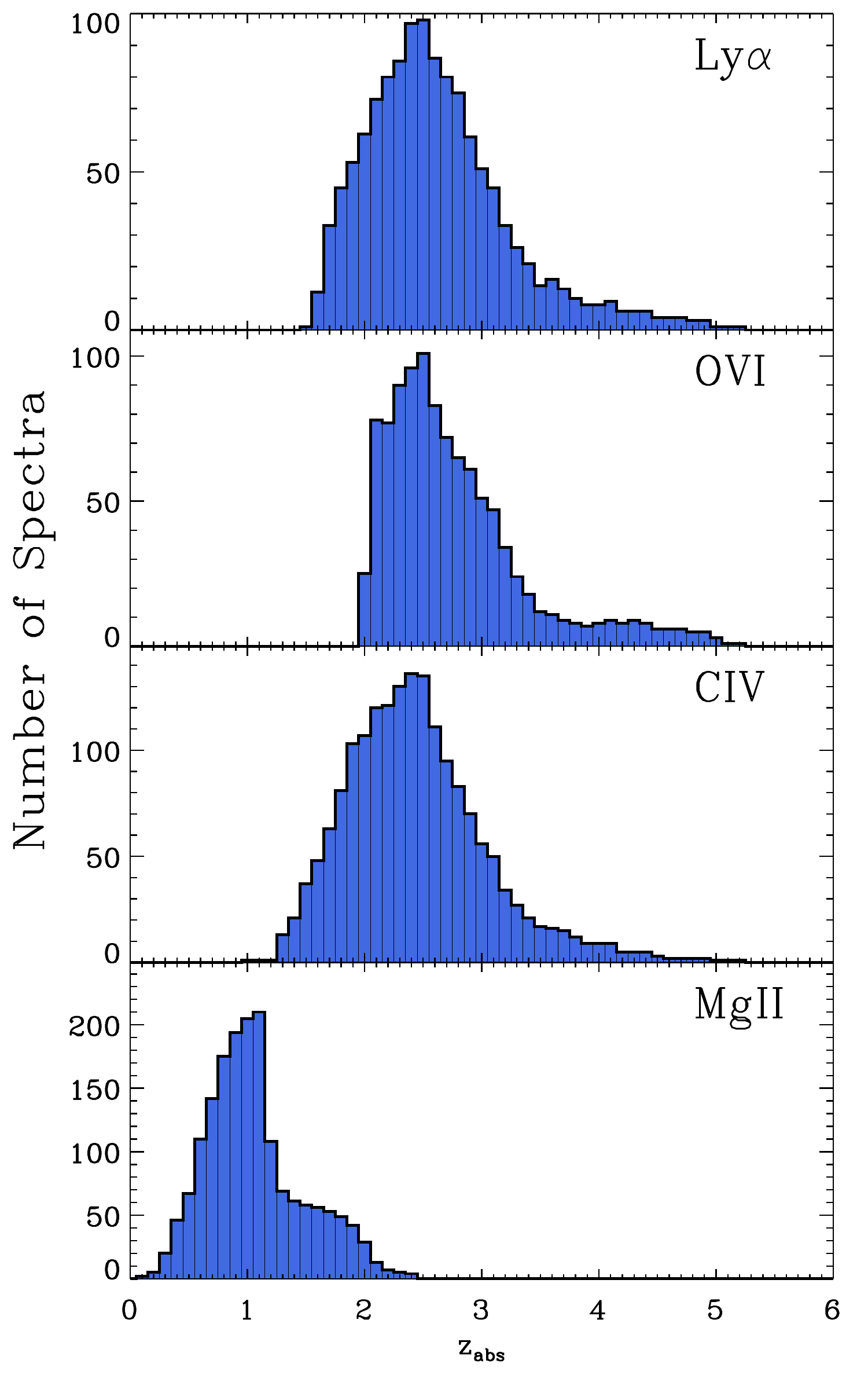}  
\caption{Redshift coverage of specific ions in the \nspectra\ spectra  of KODIAQ DR1.\label{fig_allrange}}
\end{figure*}

We anticipate that the data made available in the DR1 and future releases
will have a multitude of uses beyond those presented in
\cite{lehner14}.  To give a sense of the size of the DR1 in terms of
possible cosmological applications, we present several quantities
summarizing this rich dataset.  In Figure \ref{fig_resthist}, we
display a histogram of the wavelengths covered in the \nspectra\ of
DR1, placed in the quasar rest frame to best
display what ionic species could be
searched for in the data.  Figure \ref{fig_allrange} shows how many
spectra in DR1 provide coverage of a number of ions common in quasar
absorption line studies, namely \lya, \ovi, \civ, and \mgii.  For
\lya, we restrict ourselves to sampling only the wavelengths between the
quasar \lya\ and \lyb\ emission lines when counting sightlines.  
For \civ\ and \mgii, we sample only available wavelengths larger than that of the
quasar \lya\ line and less than the quasar \civ\ and \mgii\ emission
lines, respectively.  Finally, for \ovi, we consider all available wavelengths
lower than that of the quasar \ovi\ emission line.  

\begin{figure*}[ht]
\plotone{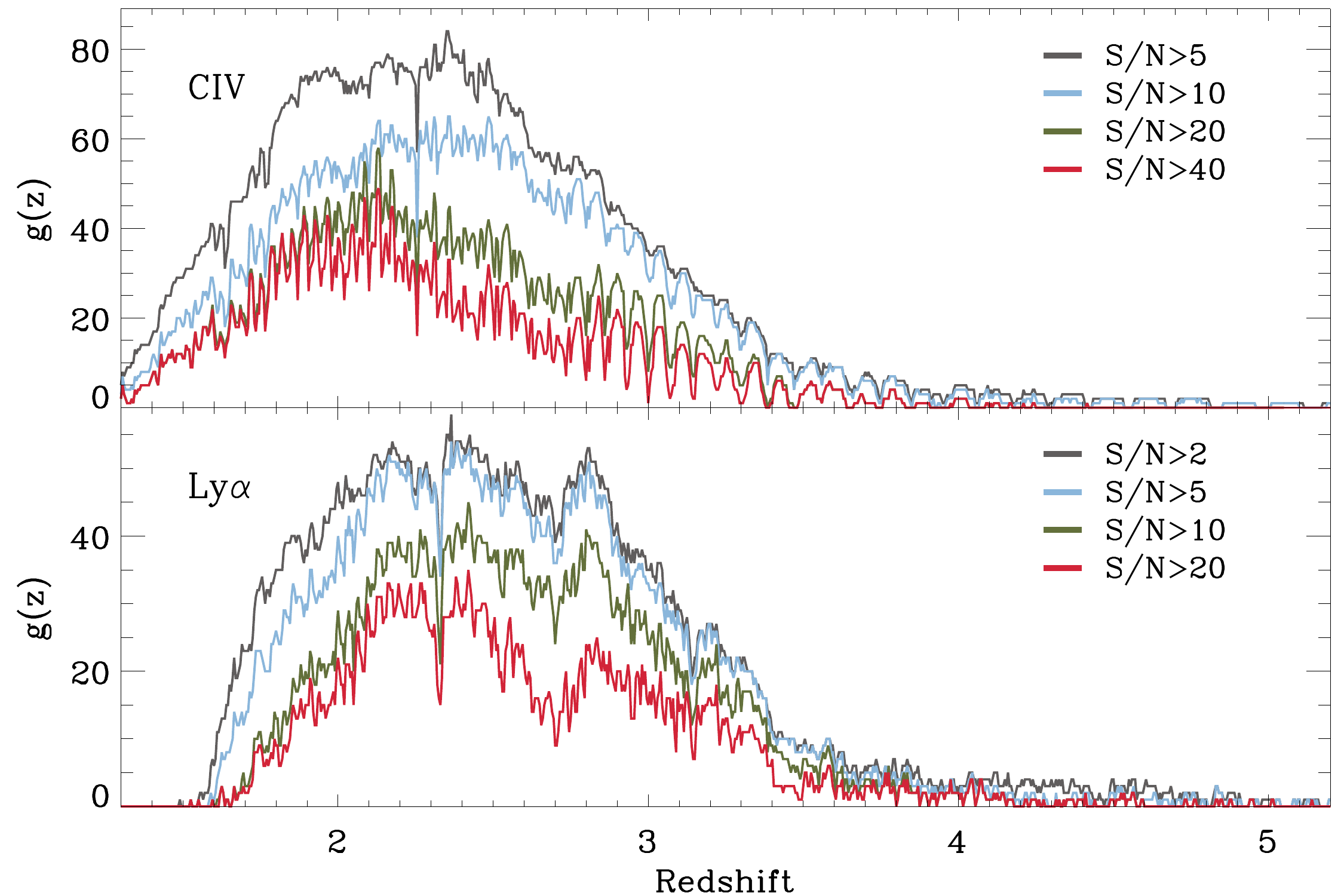}  
\caption{Redshift sensitivity function $g(z)$ for the \civt\ and
  \lya\ of the \nqsos\ qsos in DR1 \label{fig_goz}.  The sharp
  downward spikes result from the effect of the mosaic detector gaps.  
The high frequency variations result from the effects
 of the echelle blaze function.}
\end{figure*}

We can also cast
the sample in terms of the redshift sensitivity function $g(z)$ used in
studies of the incidence frequency $\ell(z) =dN/dz$ of an absorption feature
or for the column density distribution function $f(N_x,z)$ of ion $x$.
In Figure \ref{fig_goz}, we present two estimates of $g(z)$.  First,
for each of the \nqsos\ quasars in the sample (this time choosing only
a single spectrum for each quasar as to avoid double counting)
we estimate the sensitivity to searching for \lyaf\ absorption by determining which quasars
cover the \lyaf\ between 3000 \kms\ to the red of the quasar \lyb\ line
and 3000 \kms\ to the blue of the quasar \lya\ line, so as to avoid
sampling the proximity regions of the quasars themselves.  We then
calculate $g(z)$ for this sample subject to four different S/N 
critera, namely S/N\,$> 2, 5, 10,$ and 20, where the S/N is
estimated by taking the median S/N (per pixel) over a window of 150 pixels.
Second, we perform the same process, but for C~IV absorption.  For
\civ, we maintain our proximity zone exclusions, (this time between the 
\lya\ and \civ\ emission lines), but change our S/N criteria to be S/N\,$>5,10,20,$ and 40, so as to best demonstrate the high quality of the
DR1 data. Both $g(z)$ estimates show sharp downward spikes (e.g., at
$z\sim 2.3$ for \lya).  These arise due to the wavelength gaps between
the CCDs in the HIRES mosaic.  In the case of \civ, additional ripples
are present in $g(z)$.  These arise from the blaze sensitivity of the
echelle.  Figure \ref{fig_goz} demonstrates that DR1 offers a
significant dataset for any number of studies of the intergalactic and
circumgalactic medium over a significant swath of cosmic time.

\section{Summary and Future}\label{future}
We have presented here and made available to the public at the KOA the
first data release, DR1, of the KODIAQ survey.  DR1 is comprised of
\nspectra\ spectra of \nqsos\ quasars obtained with HIRES between 2004 and
2012.  The data vary in signal to noise and resolution, but a
significant portion of the spectra are of sufficient quality as to
facilitate a number of precision studies of the inter- and
circum-galactic medium  at $z\sim 2.5$.  

Additional data releases for KODIAQ are planned for the future.  DR2 will
incorporate pre-2004 spectra into the sample for another $\sim 150$
quasars in addition to augmenting some spectra in DR1.  Finally, DR3
will make available the full KODIAQ sample of all HIRES spectra in the
KOA that could be processed.  We anticipate that DR2 will be released
in 2017, and DR3 in early 2018.  All data releases will be housed at
the KOA, in addition to minor updates between data releases.

When using data products from DR1, in addition to the standard KOA acknowledgement 
(including acknowledgement to the original PIs of each program), 
we request that the community please include the following acknowledgement: 
``{\it Some/all the data presented in this work were obtained from the 
Keck Observatory Database of Ionized Absorbers toward QSOs (KODIAQ), 
which was funded through NASA ADAP grant NNX10AE84G},'' 
and cite this paper and \citet{lehner14}. 

\acknowledgements
Support for this research was made by NASA through the Astrophysics Data Analysis Program (ADAP) grant NNX10AE84G.
This research has made use of the Keck Observatory Archive (KOA),
which is operated by the W. M. Keck Observatory and the NASA Exoplanet Science 
Institute (NExScI), under contract with the National Aeronautics and 
Space Administration.   The data presented herein were obtained at the
W.M. Keck Observatory, 
which is operated as a scientific partnership among the California
Institute of Technology, 
the University of California and the National Aeronautics and Space
Administration. 
The Observatory was made possible by the generous financial support of
the W.M. Keck Foundation.
The authors wish to recognize and acknowledge the very significant
cultural role and reverence that the summit of Mauna Kea has always had within 
the indigenous Hawaiian community.  We are most fortunate to have the 
opportunity to conduct observations from this mountain.
Finally, the authors recognize the major contributions by the late 
astronomers Wal Sargent and Arthur Wolfe to 
this dataset, and indeed to the entire discipline of quasar
absorption line studies.  We dedicate this work to our memories of
these great scientists.

\bibliographystyle{apj}

\begin{deluxetable}{ccc}
\tablewidth{0pc}
\tablecaption{HIRES Deckers used in KODIAQ DR1}

\tablehead{
\colhead{Decker} &
\colhead{Width} &
\colhead{Spectra Resolution}\\
 \colhead{}& \colhead{(arcsec)} & \colhead{(FWHM)}
}

\startdata
B2 & 0.574 & $72,000$\\
B5 & 0.861 & $48,000$\\
C1 & 0.861 & $48,000$\\
C5 & 1.148 & $36,000$\\
E3 & 0.400 & $103,000$
\enddata
\end{deluxetable}

\begin{deluxetable}{llllccccl}

\tablewidth{0pc}
\tablecaption{The KODIAQ DR1 Sample}
\tabletypesize{\tiny}

\tablehead{
\colhead{Object} &
\colhead{R.A.$^a$} &
\colhead{Dec.$^a$} &
\colhead{$z_{\rm em}$$^a$} &
\colhead{Observation} &
\colhead{PI} &
\colhead{Total Exp,} &
\colhead{Decker} &
\colhead{Wavelength}
\\
\colhead{} & \colhead{(J2000)} & \colhead{(J2000)} &  \colhead{} & \colhead{Date}  & \colhead{}& \colhead{Time (s)} &\colhead{} & \colhead{Coverage (\AA)}
}

\startdata
J000931+021707 & 00:09:31.43 & +02:17:07.5 & 2.350 & December 2009 & Tytler & 1200 & C1 & 3334--6198\\
J002127--020333 & 00:21:27.35 & --02:03:33.6 & 2.596 & December 2009 & Tytler & 700 & C1 & 3366--6198\\
J002830--281704 & 00:28:30.44 & --28:17:04.8 & 2.400 & December 2009 & Tytler & 600 & C1 & 3366--6198\\
J002952+020606 & 00:29:52.12 & +02:06:06.2 & 2.333 & December 2009 & Tytler & 1000 & C1 & 3366--6198\\
J003501--091817 & 00:35:01.88 & --09:18:17.6 & 2.419 & December 2009 & Pettini & 16200 & C5 & 3123--5980\\
J004054--091526 & 00:40:54.65 & --09:15:26.8 & 4.976 & January 2011 & Wolfe & 7200 & C1 & 5453--8570\\
J004351--265128 & 00:43:51.81 & --26:51:28.0 & 2.786 & December 2009  & Tytler & 1500 & C1 & 3370--6198\\
J004358--255115 & 00:43:58.80 & --25:51:15.5 & 2.501 & December 2009  &Tytler & 900 & C1 & 3365--6198\\
J004448+372114 & 00:44:48.32 & +37:21:14.8 & 2.410 & December 2009  & Tytler & 900 & C1 & 3368--6198\\
J004530--261709 & 00:45:30.50 & --26:17:09.0 & 3.440 & August 2007 & Crighton & 14400 & C1 & 3973-8543\\
J005202+010129 & 00:52:02.40 & +01:01:29.3 & 2.271 & December 2009  &Tytler & 1000 & C1 & 3366--6198\\
J005700+143737 & 00:57:00.19 & +14:37:37.8 & 2.643 & December 2009  &Tytler & 1100 & C1 & 3366--6198\\
J005814+011530 & 00:58:14.32 & +01:15:30.2 & 2.494 & October 2004 &Wolfe & 3600 & C1 & 3386--6250\\
\nodata&\nodata&\nodata&\nodata& December 2009  & Tytler & 1200 & C1 & 3368--6198\\
J010311+131617 & 01:03:11.26 & +13:16:17.7 & 2.721 & October 2005 & Steidel & 7200 & C5 & 3123-5978\\
J010741--263328 & 01:07:41.92 & --26:33:28.4 & 2.460 & January 2008  &Tytler & 900 & C5 & 3133--6031\\
J010806+163550 & 01:08:06.41 & +16:35:50.0 & 2.652 & December 2005 &Tytler & 22204 & C5 & 3048-5895\\
J010925--210257 & 01:09:25.19 & --21:02:57.0 & 3.226 & January 2006 & Prochaska & 5400 & C5 & 3304--6179 \\
J011150+140141 & 01:11:50.07 & +14:01:41.5 & 2.470 & January 2008  & Tytler & 900 & C5 & 3133--6031\\
J012156+144823 & 01:21:56.03 & +14:48:23.8 & 2.870 & September 2004 & Prochaska & 7200 & C1 & 3244--6138\\
J013340+040059 & 01:33:40.40 & +04:00:59.0 & 4.150 & December 2006 & Wolfe & 7200 & C1 & 4159--8742\\
J014516--094517A & 01:45:16.60 & --09:45:17.0 & 2.719 & October 2005 & Steidel & 7200 & C5 & 3102--5978\\
\nodata&\nodata&\nodata&\nodata& September 2008 & Steidel & 4000 & C1 & 3022--5897\\
J014516--094517B & 01:45:16.70 &  --09:45:18 & 2.719 & October 2005 & Steidel & 17400 & C1 & 3102--5979\\
\nodata&\nodata&\nodata&\nodata& September 2008 & Steidel & 7200 & C1 & 3022--5897\\
J014944+150106 & 01:49:44.43 & +15:01:06.7 & 2.060 & December 2009  & Tytler & 1200 & C1 & 3176--6038\\
J015227--200107 & 01:52:27.34& --20:01:07.1 & 2.060 & January 2008  & Tytler & 840 & C1 & 3177--6032\\
J020455+364917 & 02:04:55.59 & +36:49:17.9 & 2.912 & October 2004 & Wolfe & 3600 & C1 & 3157--6031\\
\nodata&\nodata&\nodata&\nodata& August 2006 & Ellison & 5400 & C1 & 3473--6384\\
\nodata&\nodata&\nodata&\nodata& August 2006 & Prochaska & 2300 & C1 & 3501--6296\\
J020950--000506 & 02:09:50.70 & --00:05:06.4 & 2.828 &  September 2004 & Prochaska & 13500 & C1 & 3151-5994\\
\nodata&\nodata&\nodata&\nodata& October 2004 & Wolfe & 3600 & C1 & 4650--9219\\
\nodata&\nodata&\nodata&\nodata& September 2008 & Steidel & 9830 & C1 & 3022--5897\\
J020944+051714 & 02:09:44.62 & +05:17:14.1 & 4.180 & December 2006 & Wolfe & 7200 & C1 & 5733--7314\\
\nodata&\nodata&\nodata&\nodata& September 2007 & Wolfe & 11100 & C1 & 5211--8319\\
J021129+124110 & 02:11:29.16 & +12:41:10.8 & 2.953 & January 2011 & Wolfe & 10200 & C1 & 3681--6588\\
J022554+005451 & 02:25:54.85 & +00:54:51.9 & 2.975 & December 2006 & Wolfe & 10800 & C1 & 3571--6485\\
J022853--033737 & 02:28:53.21 & --03:37:37.1 & 2.066 & December 2009  & Tytler & 800 & C1 & 3177--6039\\
J023145+132254 & 02:31:45.89 & +13:22:54.7 & 2.059 & December 2009  &Tytler & 1400 & C1 & 3176--6039\\
J023359+004938 & 02:33:59.72 & +00:49:38.4 & 2.522 & January 2008  &Tytler & 1080 & C1 & 3174--6033\\
J023924--090138 & 02:39:24.48 & --09:01:38.6 & 2.471 & January 2008  &Tytler & 1080 & C1 & 3133--6032\\
J025127+341442 & 02:51:27.74 & +34:14:42.0 & 2.230 & January 2008  &Tytler & 780 & C1 & 3133--6031\\
J025515+014828 & 02:55:15.21 & +01:48:28.7 & 2.470 & January 2008  &Tytler & 840 & C1 & 3157--6030\\
J025644+001246 & 02:56:44.69 & +00:12:46.0 & 2.251 & January 2008  &Tytler & 900 & C1 & 3157--6030\\
J030046+022245 & 03:00:46.02 & +02:22:45.2 & 2.520 & January 2008  &Tytler & 900 & C1 & 3133--6033\\
J030341--002321 & 03:03:41.04 & --00:23:21.8 & 3.176 & September 2004 &Prochaska & 7200 & C1 & 3536--6423\\
J031003-004645 & 03:10:03.01 & --00:46:45.7 & 2.114 & January 2008  &Tytler & 1200 & C1 & 3133--6032\\
J033900--013318 & 03:39:00.90 & --01:33:18.0 & 3.197 & October 2005 & Prochaska & 9000 & C5 & 3686--6599\\
\nodata&\nodata&\nodata&\nodata& January 2006 & Prochaska & 7200 & C5 & 4185--7198\\
J040241--064124 & 04:02:41.42 & --06:41:37.9 & 2.432 & January 2011 & Wolfe & 3600 & C5 & 2995--5882\\
J045213--164012 & 04:52:13.60 & --16:40:12.0 & 2.684 & October 2005 & Steidel & 13800 & C5 & 3124--5979\\
\nodata&\nodata&\nodata&\nodata& December 2009 & Pettini & 3600 & C5 & 3100-5979\\
J073149+285448 & 07:31:49.50 & +28:54:48.6 & 3.676 & January 2006 & Prochaska & 7200 & C5 & 4160--8768\\
J074110+311200 & 07:41:10.70 & +31:12:00.2 & 0.631 & December 2006 & Wolfe & 1800 & E3 & 3076--5934\\
J074521+473436 & 07:45:21.78 & +47:34:36.2 & 3.220 & October 2007 & Milutinovic & 4500 & C1 & 3627--5535\\
J074749+443417 & 07:47:49.74 & +44:34:17.0 & 4.435 & January 2011 & Wolfe & 10800 & C1 & 4908--7983\\
\nodata&\nodata&\nodata&\nodata& July 2011 & Wolfe & 3600 & C1 & 4850--9459\\
J074927+415242 & 07:49:27.90 & +41:52:42.3 & 3.111 & March 2007 & Hamann & 12600 & C1 & 3571--8127\\
J080518+614423 & 08:05:18.18 & +61:44:23.70 & 3.033 & January 2006 & Prochaska & 5400 & C5 & 3987--8534\\
J081240+320808 & 08:12:40.68 & +32:08:08.58 & 2.712 & April 2007 & Prochaska & 14400 & C1 & 3331--6190\\
J081435+502946 & 08:14:35.19 & +50:29:46.30 & 3.897 & December 2006 & Wolfe & 10800 & C1 & 4024--8574\\
\nodata&\nodata&\nodata&\nodata& March 2008 & Wolfe & 10800 & C1 & 5219--9725\\
J081740+135134 & 08:17:40.52 & +13:51:34.5 & 4.389 & January 2011 & Wolfe & 7200 & C1 & 4935--8006\\
J082107+310751 & 08:21:07.61 & +31:07:51.2 & 2.626 & April 2007 & Sargent & 10800 & C1 & 3101--5981\\
\nodata&\nodata&\nodata&\nodata& September 2008 & Steidel & 5620 & C5 & 3058--5897\\
J082540+354414 & 08:25:40.12 & +35:44:14.2 & 3.846 & March 2008 & Wolfe & 10800 & B2 & 5180--9716\\
J082619+314848 & 08:26:19.70 & +31:48:48.0 & 3.093 & Dcember 2006 & Wolfe & 7900 & C1 & 3842--8362\\
J082849+085854 & 08:28:49.16 & +08:58:54.8 & 2.271 & April 2012 & Wolfe & 1293 & C1 & 3153--5992\\
J083052+241059 & 08:30:52.08 & +24:10:59.8 & 0.942 & December 2006 & Wolfe & 1800 & E3 & 3051--5936\\
J083102+335803 & 08:31:02.55 & +33:58:03.1 & 2.427 & December 2009 & Pettini & 18600 & C5 & 3123--5979\\
J090033+421546 & 09:00:33.49 & +42:15:46.8 & 3.290 & March 2005 & Prochaska & 6754 & C5 & 3545--8085\\
J092759+154321 & 09:27:59.78 & +15:43:21.8 & 1.805 & January 2011 & Wolfe & 12800 & C1 & 3074--5967\\
J092708+582319 & 09:27:08.88 & +58:23:19.4 & 1.910 & January 2011 & Wolfe & 21600 & C1 & 2995--5880\\
J092914+282529 & 09:29:14.49 & +28:25:29.1 & 3.399 & April 2007 & Wolfe & 10800 & C1 & 4036--7424\\
J093643+292713 & 09:36:43.51 & +29:27:13.6 & 2.924 & February 2010 & Prochaska & 7200 & C1 & 2995--5869\\ 
J094202+042244 & 09:42:02.04 & +04:22:44.6 & 3.275 & March 2005 & Wolfe & 7200 & C1 & 3364--6200\\
J095309+523029 & 09:53:09.05 & +52:30:29.7 & 1.8756 & March 2005 & Wolfe & 7200 & C1 & 3156--6039\\
J095822+014524 & 09:58:22.19 & +01:45:24.2 & 1.960 & January 2012 & Wolfe & 3600 & C1 & 3169--6030\\
J095820+322402 & 09:58:20.95 & +32:24:02.2 & 0.530 & December 2006 & Wolfe & 600 & E3 & 3078--5934\\
J095852+120245 & 09:58:52.19 & +12:02:45.0 & 3.298 & January 2006 & Prochaska & 7200 & C5 & 3608--6495\\
\nodata&\nodata&\nodata&\nodata& April 2006 & Prochaska & 1800 & C1 & 5444--8564\\
J100841+362319 & 10:08:41.22 & +36:23:19.3 & 3.126 & March 2007 & Hamann & 10800 & C1 & 3571--7993\\
J101155+294141 & 10:11:55.60 & +29:41:41.6 & 2.652 & April 2005 & Sargent & 12000 & C1 & 3101--5908\\
\nodata&\nodata&\nodata&\nodata& May 2005 & Steidel & 7200 & C5 & 3101--5983\\
\nodata&\nodata&\nodata&\nodata& December 2005 & Tytler & 12615 & C5 & 3048--5896\\
J101336+561536 & 10:13:36.37 & +56:15:36.4 & 3.633 & January 2006 & Prochaska & 3600 & C1 & 3899--8368\\
J101447+430030 & 10:14:47.19 & +43:00:30.0 & 3.125 & April 2005 & Prochaska & 5100 & C5 & 3500--6401\\
\nodata&\nodata&\nodata&\nodata& April 2007 & Wolfe & 7200 & B2 & 3975--7057\\
J101723--204658 & 10:17:23.98 & --20:46:58.6 & 2.545 & April 2007 & Sargent & 14400 & C1 & 3128--5981\\
J101939+524627 & 10:19:39.15 & +52:46:27.8 & 2.170 & April 2007 & Prochaska & 7200 & C1 & 3118--5974\\
J102009+104002 & 10:20:09.99 & +10:40:02.7 & 3.168 & April 2006 & Prochaska & 7200 & C5 & 3616--6524\\
J102325+514251 & 10:23:25.32 & +51:42:51.1 & 3.447 & March 2007 & Hamann & 6200 & C1 & 3584--7975\\
J102410+060013 & 10:24:10.44 & +06:00:13.7 & 2.131 & April 2012 & Wolfe & 3300 & C1 & 3128--5992\\
J103514+544040 & 10:35:14.22 & +54:40:40.1 & 2.989 & March 2008 & Wolfe & 10800 & C1 & 3299--7801\\
\nodata&\nodata&\nodata&\nodata& January 2008 & Prochaska & 3600 & C1 & 3368-6262\\
J104018+572448 & 10:40:18.52 & +57:24:48.1 & 3.409 & January 2006 & Prochaska & 8100 & C5 & 3794--6657\\
J104213+062853 & 10:42:13.52 & +06:28:53.0 & 2.035 & April 2012 & Wolfe & 2400 & C1 & 3128--5992\\
J105123+354534 & 10:51:23.03 & +35:45:34.3 & 4.912 & April 2010 & Wolfe & 7800 & C5 & 5127--9713\\
J105648+120826 & 10:56:48.69 & +12:08:26.8 & 1.923 & April 2006 & Prochaska & 3600 & C1 & 3101-5975\\
\nodata&\nodata&\nodata&\nodata& June 2006 & Prochaska & 7200 & C1 & 3074--5966\\
\nodata&\nodata&\nodata&\nodata& January 2011  & Wolfe & 10500 & C1 & 3074--5966\\
J110045+112239 & 11:00:45.23 & +11:22:39.1 & 4.707 & January 2011 & Wolfe & 17400 & C1 & 5362--8391\\
J110621+104432 & 11:06:21.43 & +10:44:32.6 & 1.859 & April 2012 & Wolfe & 2700 & C1 & 3325--6184\\
J110610+640009 & 11:06:10.74 & +64:00:09.6 & 2.203 & April 2007 & Sargent & 4800 & C1 & 3128--5981\\
J111113--080402 & 11:11:13.6 & --08:04:02.0 & 3.922 & April 2006 & Prochaska & 7200 & C1 & 4158--8751\\
J111909+211917 & 11:19:06.71 & +21:18:50.3 & 1.183 & June 2006 & Prochaska & 1800 & E3 & 3051--5932\\
J112442--170517 & 11:24:42.87 & --17:05:17.5 & 2.400 & May 2005 & Sargent & 12000 & C1 & 3127--5980\\
\nodata&\nodata&\nodata&\nodata& April 2007 & Sargent & 3600 & C1 & 3126--5981\\
J113130+604420 & 11:31:30.40 & +60:44:20.6 & 2.921 & December 2006 & Wolfe & 7200 & C1 & 3597--6423\\
J113418+574204 & 11:34:18.96 & +57:42:04.7 & 3.522 & January 2006 & Prochaska & 6300 & C5 & 3973--8534\\
J113508+222715 & 11:35:08.09 & +22:27:15.6 & 2.886 & December 2006 & Wolfe & 6900 & C1 & 3433--6311\\
J115538+053050 & 11:55:38.60 & +05:30:50.5 & 3.475 & April 2005 & Prochaska & 7200 & C1 & 5057--8140\\
J115940--003203 & 11:59:40.79 & --00:32:03.5 & 2.035 & April 2012 & Wolfe & 1200 & C1 & 3154--5992\\
J120207+323538 & 12:02:07.78 & +32:35:38.8 & 5.292 & April 2010 & Wolfe & 14400 & C1 & 5123--9709\\
J120416+022111 & 12:04:16.68 & +02:21:11.0 & 2.529 & April 2005 & Prochaska & 5400 & C5 & 3455--6309\\
\nodata&\nodata&\nodata&\nodata& May 2005 & Prochaska & 5400 & C5 & 3568--6422\\
J120917+113830 & 12:09:17.93 & +11:38:30.3 & 3.105 & April 2006 & Prochaska & 7200 & C5 & 3500--6400\\
J121117+042222 & 12:11:17.59 & +04:22:22.2 & 2.526 & May 2005 & Prochaska & 9000 & C1 & 3403--6302\\
J121930+494052 & 12:19:30.77 & +49:40:52.3 & 2.633 & May 2007 & Pettini & 4000 & C1 & 3127--5980\\
\nodata&\nodata&\nodata&\nodata& May 2009 & Steidel & 8000 & C5 & 3027--5880\\
J122518+483116 & 12:25:18.64 & +48:31:16.0 & 3.090 & March 2007 & Hamann & 9000 & C1 & 3571--8128\\
J122824+312837 & 12:28:24.96 & +31:28:37.6 & 2.200 & April 2007 & Sargent & 7200 & C1 & 3101--5980\\
J123643--020420 & 12:36:43.12 & --02:04:20.9 & 1.865 & April 2012 & Wolfe & 3000 & C1 & 3303--6185\\
J123748+012606 & 12:37:48.99 & +01:26:06.9 & 3.145 & April 2012 & Prochaska & 4000 & C1 & 3370--6274\\
J124610+303131 & 12:46:10.80 & +30:31:17.0 & 2.560 & June 2006 & Prochaska & 3600 & C1 & 3189--6086\\
J124924--023339 & 12:49:24.86 & --02:33:39.7 & 2.117 & January 2006 & Prochaska & 4800 & C1 & 3075--5966\\
J130426+120245 & 13:04:26.15 & +12:02:45.5 & 2.980 & March 2008 & Wolfe & 9000 & B2 & 3606--8171\\
J130411+295348 & 13:04:11.98 & +29:53:48.8 & 2.850 & February 2010 & Prochaska & 7200 & C1 & 2996--5871\\
J130542+092427 & 13:05:42.77 & +09:24:27.7 & 2.063 & April 2012 & Wolfe & 5400 & C1 & 3128--5992\\
J131040+542449 & 13:10:40.24 & +54:24:49.6 & 1.929 & March 2005 & Wolfe & 10800 & C1 & 3244--6136\\
J131215+423900 & 13:12:15.23 & +42:39:00.8 & 2.567 & April 2007 & Sargent & 9687 & C1 & 3102--5983\\
J131341+144140 & 13:13:41.19 & +14:41:40.5 & 1.884 & June 2006 & Prochaska & 3600 & C1 & 3244--6130\\
\nodata&\nodata&\nodata&\nodata& January 2011 & Wolfe & 7200 & C1 & 3243--6132\\
J131855+531207 & 13:18:55.75 & +53:12:07.2 & 2.321 & January 2006 & Prochaska & 3600 & C5 & 3838--6899\\
J133532+082404 & 13:35:32.65 & +08:24:04.2 & 1.909 & April 2012 & Wolfe & 3000 & C1 & 3326--6185\\
J134328+572147 & 13:43:28.73 & +57:21:47.2 & 3.034 & April 2007 & Prochaska & 3600 & C1 & 3075--5966\\
\nodata&\nodata&\nodata&\nodata& April 2007 & Prochaska & 3600 & C1 & 3333--6190\\
J134544+262506 & 13:45:44.50 & +26:25:06.0 & 2.031 & April 2005 & Prochaska & 4800 & C5 & 3501--6386\\
J135038--251216 & 13:50:38.88 & --25:12:16.8 & 2.534 & April 2005 & Sargent & 6000 & C1 & 3128--5981\\
J135317+532825 & 13:53:17.10 & +53:28:25.5 & 2.920 & March 2008 & Wolfe & 8400 & C1 & 3497--7985\\
J140501+444800 & 14:05:01.93 & +44:47:59.8 & 2.218 & May 2005 & Martin & 14400 & C5 & 3252--6095\\
\nodata&\nodata&\nodata&\nodata& May 2005 & Ellison & 3200 & C5 & 3215--6095\\
J141719+413237 & 14:17:19.23 & +41:32:37.0 & 2.024 & March 2005 & Wolfe & 14400 & C5 & 3304--6188\\
\nodata&\nodata&\nodata&\nodata& June 2006 & Prochaska & 10800 & C5 & 3281--6183\\
J141906+592312 & 14:19:06.31 & +59:23:12.2 & 2.321 & April 2007 & Prochaska & 7200 & C1 & 3108--5966\\
J143316+313126 & 14:33:16.10 & +31:31:26.3 & 2.940 & April 2006 & Prochaska & 6000 & C5 & 3245--6095\\
J143500+535953 & 14:35:00.54 & +53:59:53.7 & 2.635 & May 2005 & Prochaska & 7200 & C1 & 3472--6379\\
J143916--015627 & 14:39:16.27 & --01:56:27.6 & 2.162 & April 2007 & Prochaska & 5400 & C1 & 3075--5967\\
J144453+291905 & 14:44:53.54 & +29:19:05.5 & 2.660 & April 2007 & Sargent & 14400 & C1 & 3102--5981\\
\nodata&\nodata&\nodata&\nodata& May 2009  & Steidel  & 10000 & C5 & 2995--5880\\
J145435+094100 & 14:54:35.18 & +09:41:00.0 & 1.946 & April 2012 & Wolfe & 2400 & C1 & 3152--5992\\
J145408+511443 & 14:54:08.95 & +51:14:43.7 & 3.644 & July 2005 & Wolfe & 1800 & C5 & 4075--8565\\
J150932+111313 & 15:09:32.11 & +11:13:13.6 & 2.110 & April 2012 & Wolfe & 5194 & C1 & 3128--5992\\
J151224+465233 & 15:12:24.18 & +46:52:33.5 & 3.360 & May 2007 & Pettini & 11702 & C5 & 3588--6423\\
J152156+520238 &  15:21:56.48 & +52:02:38.4 & 2.208 & April 2007 & Sargent & 7200 & C1 & 3102--5982\\
J154153+315329 & 15:41:53.46 & +31:53:29.4 & 2.558 & March 2008 & Wolfe & 4800 & C1 & 4417--7480\\
J155152+191104 & 15:51:52.48 & +19:11:04.2 & 2.843 & April 2007 & Sargent & 900 & C1 & 3102--5982\\
\nodata&\nodata&\nodata&\nodata& May 2009 & Steidel & 6000 & C5 & 2995--5880\\
J155556+480015 & 15:55:56.89 & +48:00:15.0 & 3.299 & July 2005 & Wolfe & 16200 & C5 & 4078--8565\\
\nodata&\nodata&\nodata&\nodata& June 2006 & Prochaska & 10800 & C5 & 4018--8564\\
J155810--003120 & 15:58:10.16 & --00:31:20.0 & 2.827 & April 2006 & Prochaska & 4100 & C5 & 3361--6201\\
\nodata&\nodata&\nodata&\nodata& June 2006 & Prochaska & 10800 & C5 & 3335--6201\\
J155814+405337 & 15:58:14.51 & +40:53:37.0 & 2.635 & June 2006 & Chaffee & 25200 & C5 & 3101--5979\\
J160455+381214 & 16:04:55.40 & +38:12:01.0 & 2.551 & April 2005 & Sargent & 12000 & C1 & 3125--5981\\
\nodata&\nodata&\nodata&\nodata& May 2009  & Steidel & 4200 & C5 & 3030--5879\\
J160413+395121 & 16:04:13.97 & +39:51:21.9 & 3.130 & April 2007 & Wolfe & 9300 & C1 & 4291--8767\\
J160547+511330 & 16:05:47.58 & +51:13:30.2 & 1.785 & April 2007 & Prochaska & 7200 & C1 & 3118--5975\\
J160843+071508 & 16:08:43.90 & +07:15:08.6 & 2.877 & April 2007 & Prochaska & 5400 & C1 & 3078--5975\\
\nodata&\nodata&\nodata&\nodata& April 2007 & Prochaska & 5400 & E3 & 3078--5975\\
\nodata&\nodata&\nodata&\nodata& July 2008 & Murphy & 1500 & C1 & 4475--7586\\
J161009+472444 & 16:10:09.42 & +47:24:44.5 & 3.217 & April 2006 & Prochaska & 3300 & C1 & 3928--8507\\
\nodata&\nodata&\nodata&\nodata& June 2006 & Prochaska & 10800 & C5 & 3834--8356\\
J162453+375806 & 16:24:53.47 & +37:58:06.6 & 3.380 & July 2006 & Benn & 7200 & C5 & 3561--6591\\
J162548+264433 & 16:25:48.00 & +26:44:32.6 & 2.535 & May 2005 & Steidel & 12000 & C5 & 3102--5984\\
\nodata&\nodata&\nodata&\nodata& July 2006 & Ellison & 1800 & C5 & 5015--8103\\
\nodata&\nodata&\nodata&\nodata& September 2008 & Steidel & 5400 & C1 & 3022--5896\\
J162548+264658 & 16:25:48.79 & +26:46:58.7 & 2.518 & May 2005 & Steidel & 10800 & C5 & 3101--5983\\
\nodata&\nodata&\nodata&\nodata& September 2008 & Steidel & 6000 & C1 & 3022--5896\\
J162557+264448 & 16:25:57.38 & +26:44:48.2 & 2.601 & May 2005 & Steidel & 15889 & C5 & 3102--5983\\
\nodata&\nodata&\nodata&\nodata& October 2005 & Steidel & 5400 & C5 & 3124--5979\\
\nodata&\nodata&\nodata&\nodata& July 2006 & Ellison & 7200 & C5 & 5015--8123\\
J162645+642655 & 16:26:45.69 & +64:26:55.2 & 2.320 & April 2005 & Sargent & 11300 & C1 & 3127--5981\\
\nodata&\nodata&\nodata&\nodata& April 2007 & Sargent & 3840 & C5 & 3127--5981\\
J162902+091322 & 16:29:02.98 & +09:13:22.5 & 1.986 & April 2012 & Wolfe & 2400 & C1 & 3153--5992\\
J170100+641209 & 17:01:00.61 & +64:12:09.0 & 2.735 & April 2005 & Sargent & 5800 & C1 & 3129--5981\\
\nodata&\nodata&\nodata&\nodata& May 2005 & Steidel & 7800 & C5 & 3102--5985\\
\nodata&\nodata&\nodata&\nodata& July 2005 & Cowie & 4800 & B2 & 3571--8125\\
\nodata&\nodata&\nodata&\nodata& July 2005 & Cowie & 7200 & B2 & 3741--6786\\
\nodata&\nodata&\nodata&\nodata& September 2008 & Steidel & 2000 & C5 & 3060--5896\\
J171227+575507 & 17:12:27.74 & +57:55:06.8 & 3.008 & September 2004 & Prochaska & 3600 & C1 & 3471--6309\\
\nodata&\nodata&\nodata&\nodata& May 2005 & Prochaska & 3900 & C1 & 3470--6310\\
\nodata&\nodata&\nodata&\nodata& August 2006 & Prochaska & 7200 & C1 & 3451--6305\\
J173352+540030 & 17:33:52.23 & +54:00:30.5 & 3.425 & May 2005 & Prochaska & 5400 & C5 & 3735--6645\\
\nodata&\nodata&\nodata&\nodata& August 2007 & Crighton & 5400 & C1 & 3990--8543\\
J175603+574848 & 17:56:03.62 & +57:48:48.0 & 2.110 & September 2004 & Prochaska & 5400 & C1 & 3185--6047\\
\nodata&\nodata&\nodata&\nodata& October 2004 & Wolfe & 10800 & C1 & 3970--8544\\
\nodata&\nodata&\nodata&\nodata& June 2006 & Prochaska & 8800 & C1 & 3157--6048\\
\nodata&\nodata&\nodata&\nodata& August 2006 & Prochaska & 18000 & C1 & 3157--6047\\
\nodata&\nodata&\nodata&\nodata& July 2008 & Murphy & 18000 & C1 & 4111--7170\\
J182157+642036 & 18:21:57.2.0 & +64:20:36.0 & 0.297 & June 2006 & Prochaska & 2400 & E3 & 3105--5932\\
J193957--100241 & 19:39:57.25 & --10:02:41.5 & 3.787 & July 2005 & Crighton & 21600 & B5 & 3569--6618\\
\nodata&\nodata&\nodata&\nodata& August 2005 & Tytler & 18800 & C5 & 3843--6759\\
J200324--325145 & 20:03:24.11 & --32:51:45.0 & 3.783 & July 2005 & Cowie & 4800 & B2 & 3565--8300\\
\nodata&\nodata&\nodata&\nodata& July 2005 & Cowie & 7200 & B2 & 5217--8700\\
J203642--055300 & 20:36:42.29 & --05:53:00.2 & 2.582 & October 2004 & Wolfe & 10800 & C1 & 3244--6145\\
\nodata&\nodata&\nodata&\nodata& July 2005 & Wolfe & 10800 & C5 & 4078--8565\\
J210025--064146 & 21:00:25.03 & --06:41:46.0 & 3.137 & October 2005 & Prochaska & 1860 & C5 & 3722--6634\\
\nodata&\nodata&\nodata&\nodata& October 2004 & Wolfe & 18000 & C1 & 4439--8981\\
\nodata&\nodata&\nodata&\nodata& September 2007 & Wolfe & 10800 & C1 & 3889--6924\\
J212329--005052 & 21:23:29.46 & --00:50:52.9 & 2.262 & August 2006 & Prochaska & 21600 & E3 & 3027--5896\\
J212904--160249 & 21:29:04.90 & --16:02:49.0 & 2.900 & August 2007 & Crighton & 14400 & C1 & 3992--8545\\
J212912--153841 & 21:29:12.19 & --15:38:42.8 & 3.268 & September 2004 & Prochaska & 7200 & C1 & 3571--6422\\
J220639--181846 & 22:06:39.70 & --18:18:46.0 & 2.728 & September 2004 & Prochaska & 5400 & C1 & 3304--6183\\
J220852--194400 & 22:08:52.07 & --19:44:00.0 & 2.573 & September 2008 & Steidel & 12000 & C5 & 3022--5896\\
J222256--094636 & 22:22:56.11 & --09:46:36.2 & 2.926 & August 2006 & Ellison & 10800 & C1 & 3214--6096\\
J224145+122557 & 22:41:45.11 & +12:25:57.1 & 2.631 & July 2005 & Wolfe& 6300 & C5 & 4095--8551\\
J225409+244523 & 22:54:09.34 & +24:45:23.4 & 2.328 & December 2009 & Tytler & 1100 & C1 & 3366--6198\\
J230301--093930 & 23:03:01.45 & --09:39:30.7 & 3.455 & November 2005 & Wolfe & 7200 & C5 & 3837--8352\\
J231324+003444 & 23:13:24.45 & +00:34:44.5 & 2.083 & December 2009 &Tytler & 600 & C1 & 3176--6038\\
J231543+145606 & 23:15:43.56 & +14:56:06.3 & 3.390 & November 2005 & Wolfe & 4508 & C5 & 4797--7819\\
\nodata&\nodata&\nodata&\nodata& June 2004 & Prochaska & 4800 & C5 & 3926--7785\\
J233446--090812 & 23:34:46.40 & --09:08:12.3 & 3.317 & December 2006 & Wolfe & 3600 & C1 & 3761--8338\\
\nodata&\nodata&\nodata&\nodata& September 2007 & Wolfe & 10800 & C1 & 4065--7120\\
J233823+150445 & 23:38:23.16 & +15:04:45.2 & 2.420 & December 2009 &Tytler & 900 & C1 & 3367--6198\\
J234023--005327 & 23:40:23.66 & --00:53:27.0 & 2.085 & August 2006 & Prochaska & 7800 & C1 & 3048--5896\\
J234352+141014 & 23:43:52.62 & +14:10:14.6 & 2.896 & July 2005 & Chaffee & 7200 & C5 & 3296--6151\\
\nodata&\nodata&\nodata&\nodata& October 2005 & Steidel & 1800 & C5 & 3124--5977\\
\nodata&\nodata&\nodata&\nodata& December 2006 & Wolfe & 7200 & C1 & 3813--6660\\
J234451+343348 & 23:44:51.25 &+34:33:48.6 & 3.010 & October 2004 & Wolfe & 5400 & C1 & 4306--8912\\
J234628+124859 & 23:46:28.20 & +12:49:00.0 & 2.573 & October 2005 & Steidel & 7200 & C1 & 3120--5978\\
\nodata&\nodata&\nodata&\nodata& July 2008 & Murphy & 9000 & C1 & 4476--7587\\
\nodata&\nodata&\nodata&\nodata& September 2008 & Steidel & 4000 & C5 & 3022--5896\\
J234856--104131 & 23:48:56.48 & --10:41:31.2 & 3.173 & June 2006 & Chaffee & 5600 & C5 & 3102--5980\\
J235050+045507 & 23:50:50.25 & +04:55:07.8 & 2.593 & December 2009 &Tytler & 850 & C1 & 3368--6198\\
\enddata
\tablecomments{$^a$J2000 coordinates and quasar redshifts are determined by passing the quasar
  coordinates from the raw data header through SIMBAD or SDSS and
  choosing the appropriate match.}
\end{deluxetable}

\clearpage

\end{document}